\documentclass[%
 reprint,
superscriptaddress,
nofootinbib,
 amsmath,amssymb,
 aps,
]{revtex4-2}
\usepackage{caption}
\usepackage{subcaption}
\usepackage{graphicx}
\usepackage{dcolumn}
\usepackage{bm}

\begin{document}


\title{Non-Markovian paths and cycles in  NFT trades }

\author{Haaroon Yousaf}%
\affiliation{%
 Pometry \\
 \{haaroon.yousaf, ben.steer, alhamza.alnaimi\}@pometry.com
}%

\author{Naomi Arnold}
\affiliation{%
 Queen Mary University of London \\
 \{n.a.arnold, p.zhong, r.clegg\}@qmul.ac.uk
}%

\author{Renaud Lambiotte}
\author{Timothy Larock}
\affiliation{%
Oxford University \\
 \{renaud.lambiotte,larock\}@maths.ox.ac.uk
}%

\author{Richard G. Clegg}
\author{Peijie Zhong}
\affiliation{%
 Queen Mary University of London \\
 \{n.a.arnold, p.zhong, r.clegg\}@qmul.ac.uk
}%

\author{Alhamza Alnaimi}
\author{Ben Steer}
\affiliation{%
 Pometry \\
 \{haaroon.yousaf, ben.steer, alhamza.alnaimi\}@pometry.com
}%

\maketitle


Recent years have witnessed the availability of richer and richer datasets in a variety of domains, where signals often have a multi-modal nature, blending temporal, relational and semantic information. Within this context, several works have shown that standard network models are sometimes not sufficient to properly capture the complexity of real-world interacting systems. For this reason, different attempts have been made to enrich the network language, leading to the emerging field of higher-order networks \cite{lambiotte2019networks}. 
Higher-order network models can be divided into different yet related lines of research, including multi-layer networks allowing for multiple link types, combinatorial higher-order models, allowing for multi-body interactions between entities, temporal networks where the network itself is a dynamical entity, and non-Markovian higher-order networks that account for correlations in the sequence of nodes traversed by paths. 

In this work, we investigate the possibility of applying methods from higher-order networks to extract information from the online trade of Non-fungible tokens (NFTs), leveraging on their intrinsic temporal and non-Markovian nature. NFTs are digital assets such as art or memberships that are traded online between agents, often via smart contracts on a blockchain. While NFTs as a technology open up the realms for many exciting applications, its future is marred by challenges of proof of ownership, scams, wash trading and possible money laundering. We demonstrate that by investigating time-respecting non-Markovian paths exhibited by NFT trades, we provide a practical path-based approach to fraud detection.


In domains such as this, it is common to identify abnormal behaviour via anomalous structural patterns around nodes, for instance in terms of their ego network, their motifs, or the presence of suspicious cycles. Previous works \cite{nadini2021mapping} have mapped this type of data as networks, with nodes representing agents/wallets and edges their transactions. This enabled investigation into the organisation of online markets, for instance their heterogeneous and modular structure. However, the ordering of wallets possessing a given NFT, and the timestamp for each transaction is lost, despite this information being available in the data. 

This is poignant, especially when reasoning about the transfer of NFTs between counter parties, as the dynamics which drive the evolution of the system can be better understood by incorporating temporal information into the network, e.g. by imposing time-respecting constraints on consecutive edges on a path or by exploring licit vs. illicit temporal motifs. 

Tracking (fungible) money often requires the use of heuristics, e.g. a transfer of $x_{ij}$ from $i$ to $j$ at time $t$, followed by a transfer of $x_{jk}$ from $j$ to $k$ at time $t^{'}$, could be indicative of an indirect, coordinated flow between $i$ and $j$ if $x_{ij}$ and $x_{jk}$, and $t$ and $t^{'}$ are sufficiently close. In contrast, for each NFT, we can track the sequence of its transactions, allowing us to construct without ambiguity the path of an NFT among the actors. 

\begin{figure*}
     \centering
     \begin{minipage}[b]{0.50\textwidth}
         \centering
         \includegraphics[width=\textwidth]{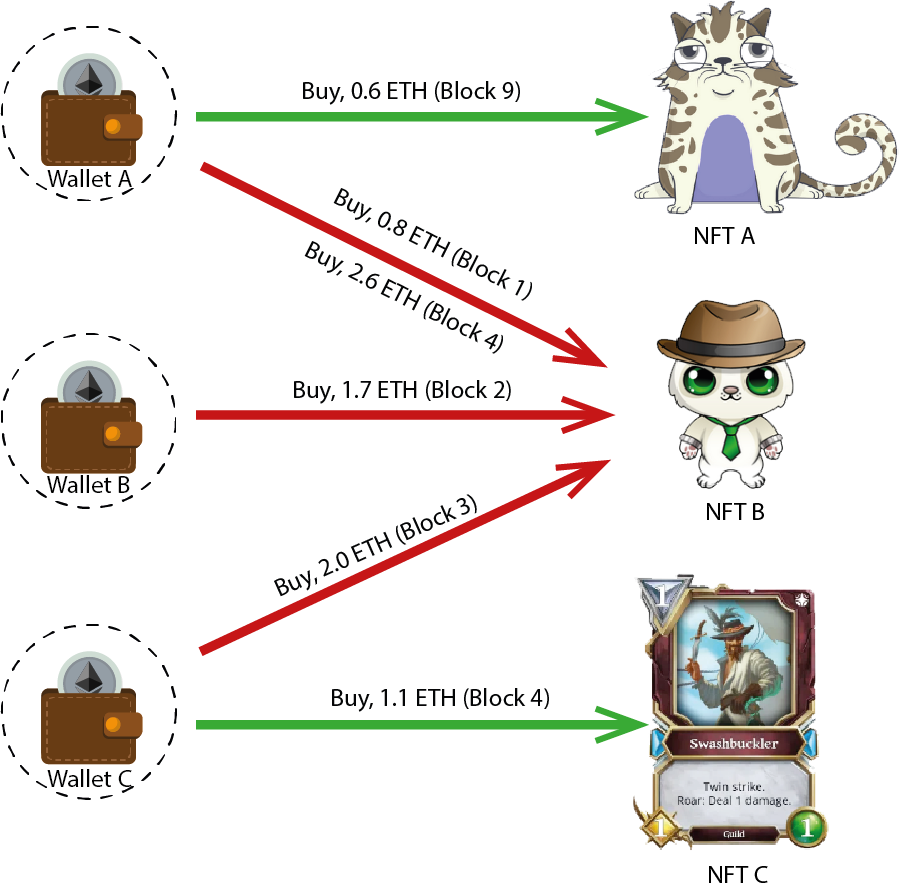}
         \subcaption{Bipartite graph representation of NFT purchases. In this artificial example, NFT $B$ is involved in the cycle of wallets $A \to B \to C \to A$.}
         \label{fig:nftschema}
     \end{minipage}
     \hfill
     \begin{minipage}[b]{0.40\textwidth}
         \centering
         \includegraphics[width=\textwidth]{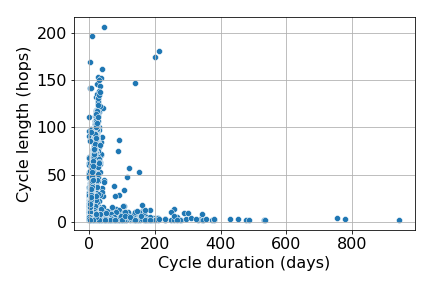}
         \subcaption{The length of each NFT cycle found in hops against the real-time duration of the cycle in days.}
         \label{fig:duration_vs_length}
         \centering
         \includegraphics[width=\textwidth]{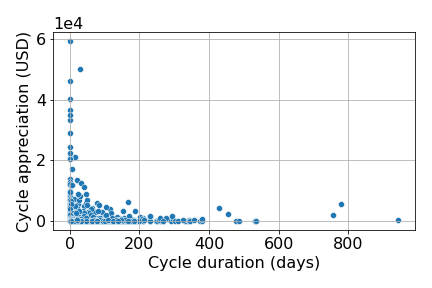}
         \subcaption{The appreciation value of each NFT involved in a cycle against the duration of the cycle in days.}
         \label{fig:duration_vs_profit}
     \end{minipage}
        \label{fig:three graphs}
\end{figure*}

Combining these premises, we may model the data as a bipartite graph consisting of wallets and NFTs, with directed edges indicating purchases. From this perspective an individual NFT can see all the wallets which have purchased it, when these occurred and for how much. With this information it is trivial to extract complex patterns, including cycles of any length. In this representation, cycles can be traced simply by scanning, for each NFT node, an ordered list of wallets which have purchased that NFT, recording a cycle if the same wallet occurs more than once in this list. Figure \ref{fig:nftschema} shows a toy example where NFT B has been involved in a cycle over the course of 4 blocks, starting and ending with wallet A in blocks 1 and 4 respectively. In contrast, extracting cycles from a transaction graph between wallets is a combinatorially difficult problem. 

To explore the efficacy of this approach we analysed all Ethereum based NFT transactions between June 23, 2017 and April 27, 2021 (2.3 million trades across 3,384 NFTs); a subset of the data from \cite{nadini2021mapping} for which the USD price was always available. This was ingested as a bipartite graph into the temporal graph engine Raphtory\cite{steer2020raphtory} where the analysis was run.


After running this algorithm, we identified 16,673 cycles whereby the user had purchased the NFT at a higher price than the original price at which they sold it. We denote this as an `appreciation'. Figures \ref{fig:duration_vs_length} and \ref{fig:duration_vs_profit} compare the cycle duration to both the length of the cycle (unique involved wallets) and appreciated price for the traded NFT. What is notable here is that the cycles involving the most wallets overwhelmingly finished in the shortest amount of time whilst generating the greatest appreciation. This is contrary to what we would expect where assets traditionally appreciate over prolonged periods of time.

From the total 352 traders completing cycles we find that the vast majority (322 traders \(\sim\) 91.48\%) had been involved in 4 or fewer. A small contingent in the middle (10 traders \(\sim\) 2.84\%) conducted between 4 and 14. This leaves a small subset of `Whales' (20 traders \(\sim\) 5.68\%) involved in between 528 and 617 cycles each. 

The wallet involved in the most cycles (617) traded 32 NFTs belonging to Etheremon, Cryptokitties and Mlbchampion between 20 wallets across 12,874 transactions, gaining a total appreciation of 1,389 USD (averaging \$2.25 per trade). By exploring the inter-purchase times of each transaction within the cycles we identified that over 88\% of transactions executed in 3.4 hour to 4.6 hour increments. Repeating this analysis on the rest of the whales we discovered that these too followed the same pattern, purchasing and trading NFTs between the same groups of wallets, with the same 3-4 hour time window. 

Due to the small price changes on these NFTs across the cycles (probably less than the total fees paid) the actions of these accounts could be explained by a number of things. They could be owned by the same entity or group performing wash-trades, a veteran NFT trader transacting within their close knit community or a case of NFT creators employing automated Market Maker bots to generate liquidity for their tokens. 

Although our initial focus was on identifying patterns of NFTs used as a vehicle for wash trading and money-laundering, we show that by incorporating time into traditional analytical network and path finding models, we can extract at scale previously unknown patterns to help unmask the purposes of these transactions. The capabilities of the Raphtory system have considerable promise in many different contexts for analysis of temporal graphs, particularly where richer information is available and/or where processing at scale is important. We intend to extend this work, investigating the implication of other paths and motifs. Further feeding into the allocation of legitimacy or risk scores to individual NFTs.


\bibliography{apssamp}

\end{document}